\documentclass[
reprint,
superscriptaddress,
amsmath,amssymb,
aps,
prb,
twocolumn,
floatfix,
]{revtex4-2}

\usepackage{xr}

\makeatletter
\newcommand*{\addFileDependency}[1]{
\typeout{(#1)}
\@addtofilelist{#1}
\IfFileExists{#1}{}{\typeout{No file #1.}}
}\makeatother

\usepackage{graphicx}
\usepackage{verbatim}
\usepackage[version=4]{mhchem}
\usepackage{xcolor}
\usepackage{physics}
\usepackage[normalem]{ulem}
\usepackage{pdfpages}
\usepackage{pgffor}
\usepackage{soul}
\usepackage{booktabs}
\usepackage{multirow}
\usepackage{hyperref}

\makeatletter
\AtBeginDocument{\let\LS@rot\@undefined}
\makeatother

\begin{document}

\title{Prediction rigidities for data-driven chemistry}

\author{Sanggyu Chong}
\affiliation{Laboratory of Computational Science and Modeling, Institute of Materials, \'Ecole Polytechnique F\'ed\'erale de Lausanne, 1015 Lausanne, Switzerland}

\author{Filippo Bigi}
\affiliation{Laboratory of Computational Science and Modeling, Institute of Materials, \'Ecole Polytechnique F\'ed\'erale de Lausanne, 1015 Lausanne, Switzerland}

\author{Federico Grasselli}
\affiliation{Laboratory of Computational Science and Modeling, Institute of Materials, \'Ecole Polytechnique F\'ed\'erale de Lausanne, 1015 Lausanne, Switzerland}

\author{Philip Loche}
\affiliation{Laboratory of Computational Science and Modeling, Institute of Materials, \'Ecole Polytechnique F\'ed\'erale de Lausanne, 1015 Lausanne, Switzerland}

\author{Matthias Kellner}
\affiliation{Laboratory of Computational Science and Modeling, Institute of Materials, \'Ecole Polytechnique F\'ed\'erale de Lausanne, 1015 Lausanne, Switzerland}

\author{Michele Ceriotti}
\email{michele.ceriotti@epfl.ch}
\affiliation{Laboratory of Computational Science and Modeling, Institute of Materials, \'Ecole Polytechnique F\'ed\'erale de Lausanne, 1015 Lausanne, Switzerland}

\begin{abstract}
The widespread application of machine learning (ML) to the chemical sciences is making it very important to understand how the ML models learn to correlate chemical structures with their properties, and what can be done to improve the training efficiency whilst guaranteeing interpretability and transferability.
In this work, we demonstrate the wide utility of prediction rigidities, a family of metrics derived from the loss function, in understanding the robustness of ML model predictions.
We show that the prediction rigidities allow the assessment of the model not only at the global level, but also on the local or the component-wise level at which the intermediate (e.g. atomic, body-ordered, or range-separated) predictions are made. 
We leverage these metrics to understand the learning behavior of different ML models, and to guide efficient dataset construction for model training.
We finally implement the formalism for a ML model targeting a coarse-grained system to demonstrate the applicability of the prediction rigidities to an even broader class of atomistic modeling problems.
\end{abstract}

\maketitle

\section{Introduction}

In data-driven chemistry, computational and experimental data~\cite{Grazulis2009, Groom2016, Westbrook2003, Jain2013, Talirz2020} is exploited to deduce new insights that are beneficial for the mechanistic understanding of chemical processes.
Data-driven chemistry relies on machine learning (ML) models,~\cite{Butler2018, Westermayr2021, Ceriotti2022} which exhibit greater flexibility and scalability to larger datasets compared to pre-existing regression methods.
One crucial aspect to consider in ML is that the models are intrinsically statistical, and hence their predictions are always made with a degree of uncertainty.~\cite{Musil2019, Imbalzano2021, Kellner2024} 
This can be exploited to understand when and when not to trust the model predictions by reliably quantifying their uncertainties.

In this application domain, ML models are often built to predict a quantity as a sum of constituent terms rather than directly predicting the global, physical observable associated with a given chemical system.
Examples include predicting the local energies of constituent atoms as opposed to the total energy of an entire system,~\cite{Behler2007, Bartok2010, Drautz2019, Schutt2021, Musaelian2023, Pozdnyakov2023} or combining predictions at multiple length-scales.~\cite{Grisafi2019, Grisafi2021, Ko2021, Huguenin2023} 
Such approaches enhance the transferability of ML models, and offer a heuristic understanding of complex chemical phenomena as projections over interpretable components of the system.~\cite{Deringer2018b, Deringer2018si, El-Machachi2022, Gardner2023}
However, they contribute a degree of arbitrariness to the ML models, as the global target properties can be decomposed in many different ways.~\cite{Eckhoff2019, Tisi2021, Pegolo2022}
Consequently, the interpretability and transferability of the ML model are also connected to the quality and robustness of these intermediate predictions.

To better understand the implications of arbitrariness in the target decomposition, some of us have recently proposed prediction rigidities as metrics to quantify the robustness of ML model predictions.~\cite{Chong2023, Bigi2024}
Prediction rigidities are derived from a constrained loss formulation to quantify the degree of sensitivity, or ``rigidity'', of a ML model when the value of one prediction is perturbed away from that obtained from the unconstrained model.
From a practical perspective, they allow for an understanding of how stable the ML model predictions are with respect to changes in the model architecture or dataset makeup.
One can easily derive several different versions of the prediction rigidity depending on where the constrained loss formulation is applied.
This allows for a form of ``introspection'' of the ML models, even at the level of intermediate (e.g. atomic, body-ordered, or range-separated) predictions.
The prediction rigidities are also versatile in that the precise details of model training, e.g. incorporation of multiple loss terms, weighting of different training samples, can be exactly accounted for.

\vspace{+0.5cm}
In this work, we demonstrate the utility of prediction rigidities in ML for chemical sciences under a wide range of atomistic modeling scenarios. First, the theory behind the prediction rigidities is briefly revisited. Next, a practical extension of the prediction rigidities to neural network-based ML models is demonstrated, which we then use to explore the learning dynamics of such models. This is followed by a section where the global and local prediction rigidities are used to guide the efficient construction of a training set, where these metrics make a difference in resolving degeneracies and decreasing the error for the systems of interest. Subsequently, we analyze the learning behavior of multi-component models (e.g.~a body-ordered model, a multi-length-scale model), showing that orthogonalization of different components can improve their interpretability. Finally, the wide applicability of the PRs is showcased by implementing the formalism for a coarse-grained ML model of water and observing that one can use the metrics to monitor convergence and to detect potential failures.
\newpage

\begin{table*}[t]
\small
  \caption{Types of prediction rigidities presented in this work, along with the purpose they serve and the corresponding $\mathbf{g}_\star$ used in their derivations.}
  \label{tbl:prs_extended}
  \begin{tabular*}{\textwidth}
  {@{\extracolsep{\fill}}lp{0.4\textwidth}p{0.3\textwidth}p{0.11\textwidth}}
    \toprule
    & Name and purpose & Prediction type & Form of $\mathbf{g}_\star$ in~\eqref{eq:pr} \\
    \midrule
    \textbf{PR} & \textbf{\textit{Prediction rigidity}} -- for assessing the confidence on the global predictions of ML models & Global prediction, $\tilde{Y}_\star$ & 
    \vspace{-.65cm}\begin{equation*}
        \left. \frac{\partial \tilde{Y}_\star}{\partial \mathbf{w}} \right|_{\mathbf{w}_o}
    \end{equation*} \\ \\
    \textbf{LPR} & \textbf{\textit{Local prediction rigidity}} -- for assessing local predictions of models that incorporate a locality ansatz & Local prediction for environment $j$, $\tilde{y}_j$ 
    \newline \begin{equation*}
        \tilde{Y}_\star = \sum_{j \in \star} \tilde{y}_j
    \end{equation*} 
    &\begin{equation*}
    \left.\frac{\partial \tilde{y}_j}{\partial  \mathbf{w}}\right|_{\mathbf{w}_o}    
    \end{equation*}
    \\ \\
    \textbf{CPR} & \textbf{\textit{Component-wise prediction rigidity}} -- for separately assessing different prediction components of models that incorporate several additive prediction components & Prediction component, $\tilde{Y}_C$ \newline \begin{equation*}
        \tilde{Y}_\star = \tilde{Y}_{C_1} +  \tilde{Y}_{C_2} + \cdots
    \end{equation*} (e.g. body-orders, multiple length-scales)
    & \begin{equation*}
        \left.\frac{\partial \tilde{Y}_C}{\partial \mathbf{w}}\right|_{\mathbf{w}_o}        
    \end{equation*}\\ 
    \bottomrule
  \end{tabular*}
\end{table*}

\section{Theory}

In this section, we present the theoretical background of prediction rigidities for atomistic ML models.
As there exist two previous publications where the general derivation of prediction rigidities was presented in detail,\cite{Chong2023, Bigi2024} here we exclusively focus on how the prediction rigidities can be formulated for ML models in the field of chemical sciences.
\subsection{Prediction rigidity (PR)}
The name ``prediction rigidity'', hereon abbreviated as PR, comes from the mathematical construction devised by Chong et al.~\cite{Chong2023} to quantify the response of a regression model, in terms of its loss, to a small perturbation $\Delta\epsilon_\star$ (where $\star$ denotes a specific sample) in the prediction that is imposed through a Lagrange multiplier. 
By taking a constrained loss minimization approach, one obtains an expression for the change in model loss with respect to the optimum:
\begin{equation}
    \Delta\mathcal{L} = \frac{1}{2} R_\star \, \Delta\epsilon_\star^2 + \mathcal{O}[\Delta\epsilon_\star^3].
\end{equation}
$\Delta\mathcal{L}$ is proportional to the square of $\Delta\epsilon_\star$, and the corresponding coefficient $R_\star$ defines the PR.
Several different types of PR can be defined by targeting different terms in the model (e.g. local, component-wise, etc.) with $\Delta\epsilon_\star$, all sharing the following structure:
\begin{equation}\label{eq:pr}
    R_\star =  \Big( \mathbf{g}_\star^\top \mathbf{H}_o^{-1} \mathbf{g}_\star \Big)^{-1},
\end{equation}
where 
\begin{equation}
    \mathbf{H}_o \equiv \left.\frac{\partial^2 \mathcal{L}}{\partial \mathbf{w}\partial \mathbf{w}^\top}\right|_{\mathbf{w}_o}\label{eq:hess}
\end{equation}
is the Hessian of the loss $\mathcal{L}$ with respect to the weights $\mathbf{w}$ computed at the optimum $\mathbf{w}_o$. Note that $\mathbf{H}_o$ does not depend on the specific sample or prediction type. Only the vector $\mathbf{g}_\star$ does, and it can be easily adjusted to target different prediction types as outlined in Table \ref{tbl:prs_extended}. Additionally, the PRs do not require the target values used for regression when the model is trained with a squared loss. From Eq. \eqref{eq:pr}, it is evident that the PR has the meaning of the inverse norm of $\mathbf{g}_\star$ using the matrix $\mathbf{H}_o^{-1}$ as the metric tensor. Note that similar expressions can be identified in the formulations of pre-existing approaches for uncertainty quantification and active learning.~\cite{Mackay1992, Rasmussen2005}

\subsection{Versatile formulation for arbitrary losses}
Given the versatility in their mathematical construction, the PRs can easily account for different loss forms. Here, we take the case of ML interatomic potentials (MLIPs) as a practical example, since it is one of the most widespread applications of ML for chemical sciences.~\cite{Behler2007, Bartok2010,Drautz2019,Schutt2021,Batatia2022MACE,Bigi2022,Bigi2023,Pozdnyakov2023}
MLIPs are trained on first-principles energies, often in conjunction with the forces and/or stresses. The loss for the model corresponds to
\begin{equation}
    \mathcal{L}(\mathbf{w}) = \sum_{i=1}^{N_{\rm{train}}} \ell_i
\end{equation}
where
\begin{equation}
\begin{aligned}
    \ell_i = {} & \lambda_E (E_i-\tilde{E}_i)^2 \\ & + \lambda_f \sum_\alpha (f_{i, \alpha}-\tilde{f}_{i, \alpha})^2 \\ & + \lambda_s \sum_\beta (s_{i, \beta}-\tilde{s}_{i, \beta})^2.
\end{aligned}
\end{equation}
Here $E$, $f$, and $s$ are energy, forces and stresses, $\alpha$ and $\beta$ correspond to different force and stress components, and $\lambda$ denotes the weight applied to each loss term. In such a case,

\begin{equation}
\begin{aligned}
    \mathbf{H} \approx {}  \sum_{i \in \mathcal{D}} \Bigg[ & \lambda_E \Big( \frac{\partial E_i}{\partial \mathbf{w}} \Big)^\top \frac{\partial E_i}{\partial \mathbf{w}} \\ & + \lambda_f \sum_\alpha \Big( \frac{\partial f_{i, \alpha}}{\partial \mathbf{w}} \Big)^\top \frac{\partial f_{i, \alpha}}{\partial \mathbf{w}} \\ & + \lambda_s \sum_\beta \Big( \frac{\partial s_{i, \beta}}{\partial \mathbf{w}} \Big)^\top \frac{\partial s_{i, \beta}}{\partial \mathbf{w}} \Bigg],
\end{aligned}
\end{equation}

\noindent which follows from using a generalized Gauss-Newton approximation of the Hessian. This avoids prohibitively expensive calculations of the second derivatives of the model. More information on this approximation and on its application to arbitrary loss functions can be found in Appendix~\ref{app:generic-loss}.

\subsection{Formulation for neural networks}\label{subsec:theory-ll}

The PR can be formulated in a simple, closed form for both linear and kernel models. In the case of ``deep'' neural network (NN) models that are widely used for their enhanced flexibility and scalability to large datasets, the formulation is less obvious.
Although the application of PRs to NN models can be attempted by treating the entire NN as a pseudo-linear model~\cite{Chong2023}, the quadratic scaling of the pseudo-Hessian matrix with respect to the number of NN parameters makes it impractical.

Recently, Bigi et al.~\cite{Bigi2024} have proposed a different approach to conveniently obtain the PRs of NN models.
Grounded on the theories of the Laplace approximation and neural tangent kernel (NTK), they properly justify the exclusive use of the last-layer latent features of the NN in calculating the PRs when the model is linear in the final readout layer for prediction. Note that similar last-layer approaches can be found in other related works.~\cite{Janet2019,Zhu2023,Kellner2024,Harrison2024}
The last-layer PR is then given by modifying~\eqref{eq:pr} so that only the weights of the last layer are considered for the derivatives in Eq.~\eqref{eq:hess}, and in the definitions of $\mathbf{g}_\star$ in Table \ref{tbl:prs_extended}. 
For instance, assuming a loss function given by a sum of squared errors, the last-layer PR is given by
\begin{equation}\label{eq:llpr}
    R_\star^{ll} \propto (\mathbf{f}_\star^\top (\mathbf{F}^\top \mathbf{F} + \varsigma^2 \mathbf{I})^{-1} \mathbf{f}_\star)^{-1},
\end{equation}
where $\mathbf{F}$ is a matrix (of dimensions $N_{\mathrm{train}} \times N_{\mathrm{features}}$) containing the last-layer latent features for every sample in the training set, $\mathbf{f}_\star$ are the analogous features for sample $\star$ under consideration, and $\varsigma^2 \mathbf{I}$ term is a regularization term. Likewise, the last-layer local prediction rigidity (LPR) is given by
\begin{equation}\label{eq:lllpr}
    R_j^{L, ll} \propto (\mathbf{f}_j^\top (\mathbf{F}^\top \mathbf{F} + \varsigma^2 \mathbf{I})^{-1} \mathbf{f}_j)^{-1},
\end{equation}
where $\mathbf{f}_j$ are the last-layer latent features for environment $j$.

In the following section, we present the application of this approach to NN-based atomistic ML models in performing uncertainty quantification and assessing their learning dynamics.
We remark that the computational cost to obtain these last-layer PRs is typically small, no re-training or modification to the NN model is needed, and that the formalism can also be applied to the trained NN models.
\begin{figure*}[t]
\centering
\includegraphics[width=15cm]{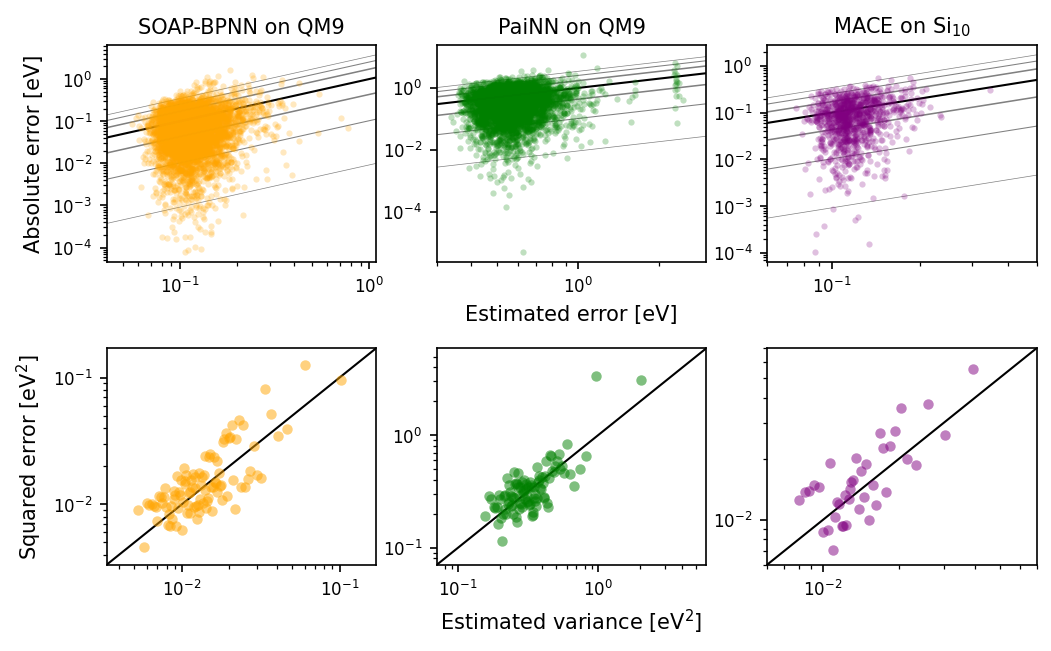}
  \caption{Trends between empirical errors and estimated model uncertainties for SOAP-BPNN, PaiNN, and MACE. The top row presents the absolute error vs. estimated error. The bottom row shows the plots of squared error vs. estimated variance (inverse of PR), where each point is an average over 50 (SOAP-BPNN and PaiNN) or 25 (MACE) test set samples with similar estimated variance values. Results for SOAP-BPNN and PaiNN are on the QM9 dataset, and for MACE are on the \ce{Si10} dataset. In all plots, $y=x$ line is shown in black. In the top row, isolines that enclose fractions of the total probability equivalent to $\sigma$, $2\sigma$ and $3\sigma$ of a Gaussian distribution (approximately 68\%, 95\%, 99\%) are shown in gray.~\cite{Bigi2024, Kellner2024}}\label{fig:nn-pr}
\end{figure*}

\begin{figure*}[t!]
\centering
\includegraphics[width=15cm]{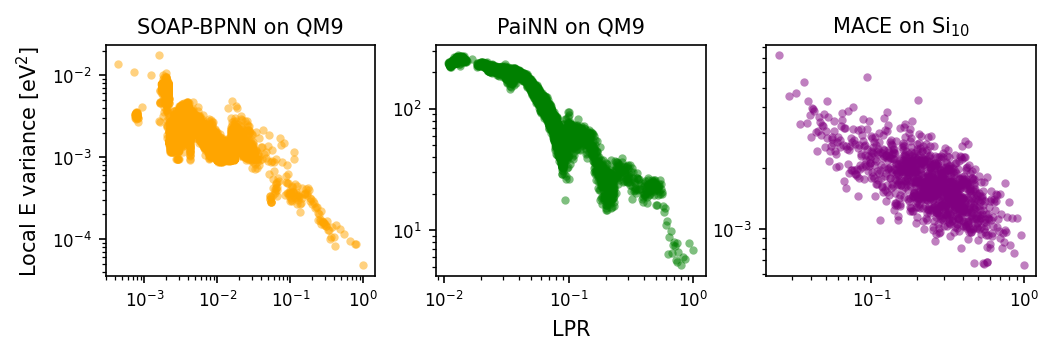}
  \caption{Variance in the local energy predictions vs. last-layer LPR for SOAP-BPNN, PaiNN, and MACE. Results for SOAP-BPNN and PaiNN are on the QM9 dataset, and for MACE are on the \ce{Si10} dataset. The LPRs are normalized to the maximum observed value from the test set in each case. Each point corresponds to an average over 100 local predictions with similar LPR values.}
  \label{fig:nn-lpr-vs-spread}
\end{figure*}

\section{PRs of NN models}

In this section, we study the last-layer PRs of three representative NN-based atomistic ML models: a Behler-Parrinello NN~\cite{Behler2007} that takes the smooth overlap of atomic positions (SOAP)~\cite{Bartok2013} as input (hereon referred to as SOAP-BPNN), a polarizable atom interaction NN (PaiNN),~\cite{Schutt2021} and MACE~\cite{Batatia2022MACE}. The first two models were trained on the QM9 dataset~\cite{Ramakrishnan2014} to predict the total energies of the molecules, using smaller subsets of 10,000 training, 1000 validation, and 5000 test samples. For MACE, a different dataset composed of \ce{Si10} clusters with 8000 training, 1000 validation, and 1000 test samples was employed, also using the total energies as targets. Full details of model training and dataset acquisition can be found in the Supplementary Information.

\subsection{Last-layer PRs of NN models}

We start by establishing the validity of a last-layer approximation when computing the PR of NN-based models. To do so, we show that the inverse of the last-layer PR can be used to quantify the model uncertainty in the total energy for the three architectures.~\cite{Bigi2024} Results in Figure \ref{fig:nn-pr} show the correlations between empirical errors of the model on the test set vs.~their estimates using the inverse of the PR. A linear correlation between the actual and estimated errors can be clearly observed for all three models, and across the entire range of consideration, which shows the validity of the last-layer PR as a metric to quantify the robustness of NN-based atomistic model predictions.

We also consider the last-layer LPR of the three NN-based models. As there exist no physical targets for the local energies, we performed ten additional training runs for each NN model on a 10-fold sub-sampled dataset, and analyzed the variance in the local predictions with respect to the committee average. Figure \ref{fig:nn-lpr-vs-spread} shows that in all three cases, a clear inverse trend between the local energy variance and the last-layer LPR exists, indicating that the local predictions for the high LPR environments are more robust, and vice versa. These results corroborate the efficacy of the last-layer approximation in also computing the LPR of NN-based atomistic ML models.

\subsection{Assessing the learning dynamics}

We now consider the impact of training details on the PR distribution. In NN-based models, the training is almost always carried out via numerical optimization, as opposed to linear or Gaussian process regressors that are commonly trained in an analytical, deterministic manner. Hence, we first investigate the changes in the model along the optimization ``trajectory'' by computing the last-layer LPR of the local environments in the test set for several intermediate checkpoints of a SOAP-BPNN model along its optimization trajectory.

\begin{figure}[t!]
\includegraphics[width=\columnwidth]{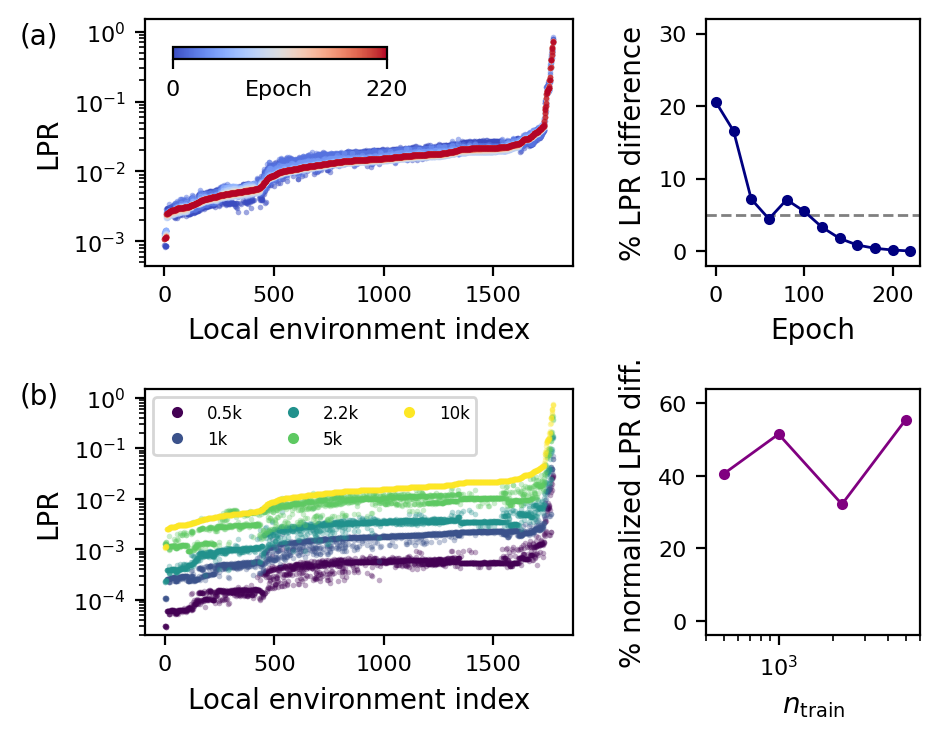}
  \caption{Dependence of the last-layer LPR distribution of 100 test set molecules on (a) model optimization trajectory, and (b) training set size, for the SOAP-BPNN model trained on QM9. The left panels show raw LPR distributions, where the local environments are ordered by the LPR distribution of the final model trained on the full dataset to numerical convergence. Right panels show the average (normalized) LPR differences of the intermediate models from the final model.}
  \label{fig:lpr-dynamics}
\end{figure}
Figure \ref{fig:lpr-dynamics}a shows that the randomly initialized model at epoch 0 already captures the overall trend in the LPRs observed in the final model. As the learning progresses, the LPR distribution quickly converges, as apparent from the lack of clearly distinguishable markers for epochs between 120 and 220. In fact, the average relative change in the LPRs with respect to the original model drops below 5\% at epoch 120, and below 1\% at epoch 160. 
We conjecture this to be connected to the Gaussian-process neural network theory in the last-layer approximation,~\cite{Jacot2018,Lee2020,Daxberger2021} whereby the NTK calculated with last-layer weights remains approximately constant during training and independent of random initial values in the limit of infinitely wide neural networks. Within this theory, the equivalence between the linear and Gaussian process formalisms implies the approximate invariance and initialization-independence of the LPR (and all other PRs) for sufficiently wide neural networks.

Next, we consider the dependence of the PR distribution on dataset size by training additional SOAP-BPNN models using smaller training subsets of 500, 1000, 2233, and 5000 structures while keeping the validation set fixed. Resulting changes in the last-layer LPR of local environments in the test set are presented in Figure \ref{fig:lpr-dynamics}b. Here, apart from the increasing trend due to the growth of the training set, notable differences in the relative LPR distributions are observed across the models, with average \emph{normalized} (by LPR$_{j}$ of each model, where $j$ is the highest LPR environment for the original model) relative LPR differences of 41\%, 52\%, 32\% and 55\% with respect to the original model. This is explained by how the smaller subsets of the training set can describe entirely different loss landscapes for the model. It also highlights the importance of judicious dataset composition in achieving a robust description for the systems of interest, which is further investigated in the next Section.

\begin{figure}[t!]
  \includegraphics[width=\columnwidth]{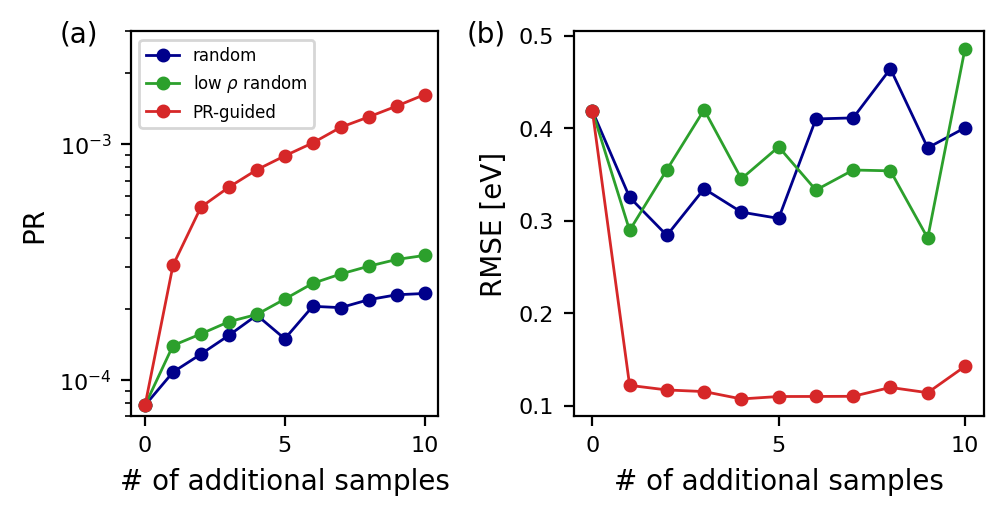}
  \caption{Results of dataset augmentation for the linear LE-ACE carbon model in extrapolating to surface-containing systems while only being trained on the bulk systems. Panel (a) shows the PR and panel (b) shows the energy RMSE of surface-containing carbon structures when up to 10 additional samples are added to the training set. Three approaches for structure selection are considered: random selection, random selection of low-density ($\rho <$ 2.0 g/cm$^3$) samples only, and the iterative PR-guided selection. Plots for the random selection cases are the average of results from 10 random seeds.}
  \label{fig:data-aug}
\end{figure}

\section{PR-guided dataset construction}
In this section, we demonstrate the utility of the PRs in guiding efficient dataset construction for ML model training. Such an effort becomes important when computational resources are limited, the desired level of theory -- and thus the computational cost of additional reference calculations -- is high, and/or it is necessary to refine a pre-trained ML model on a curated set of additional structures, as in the fine-tuning of universal or foundation models for a specialized application.~\cite{Chen2022, Deng2023, Batatia2024, Yang2024, Focassio2024, Kaur2024} We target such scenarios in two separate case studies: in the first, we consider the case of fine-tuning a trained model to extend its applicability to another system of interest. In the second, we consider how the PRs can be exploited in the active learning of atomistic ML models, where one seeks to identify the structure(s) that best resolve the uncertainty within each iteration of the active learning process.
 
\subsection{PR-guided dataset augmentation}\label{sec:data-aug}
As a case study, we consider extending the applicability of a ML model trained on the total energies of bulk carbon structures to surface-containing structures. For this, a linear Laplacian eigenstate ACE (LE-ACE) model~\cite{Bigi2022} is trained on a set of 800 bulk, high density ($\rho > 3.0$ g/cm$^3$) amorphous and liquid carbon structures taken from the GAP-17 carbon dataset.~\cite{Deringer2017} We attempt to make this model transferable to surface-containing structures by adding a few additional structures and re-training the model, where the additional structures are selected from a larger candidate pool of bulk amorphous carbon structures that span the entire density range. While there exists an obvious approach of directly incorporating the surface-containing structures, we limit the choice to bulk structures for the sake of highlighting the utility of our proposed metrics. This creates a challenging scenario where we attempt to achieve model applicability for one system by incorporating a few samples from another system. Full details of model training and target-oriented dataset augmentation are provided in the Supplementary Information.

Here we devise an iterative, PR-guided dataset augmentation strategy where, given a candidate pool, the structure that most increases the PR for the target systems is selected at each iteration. The added structure is promptly taken into account by updating $\mathbf{H}_o$ as multiple structures get selected. We demonstrate this strategy for the present case study by selecting up to 10 bulk structures that best improve the PR of the surface-containing systems. For reference, we also perform random selection from the entire candidate pool, as well as random selection of low-density structures ($\rho < 2.0$ g/cm$^3$) that are more likely to contain surface-resembling local environments as a less na\"{i}ve baseline. Both random selection approaches are repeated 10 times with different random seeds.

\begin{figure*}[t!]
\centering
  \includegraphics[width=16cm]{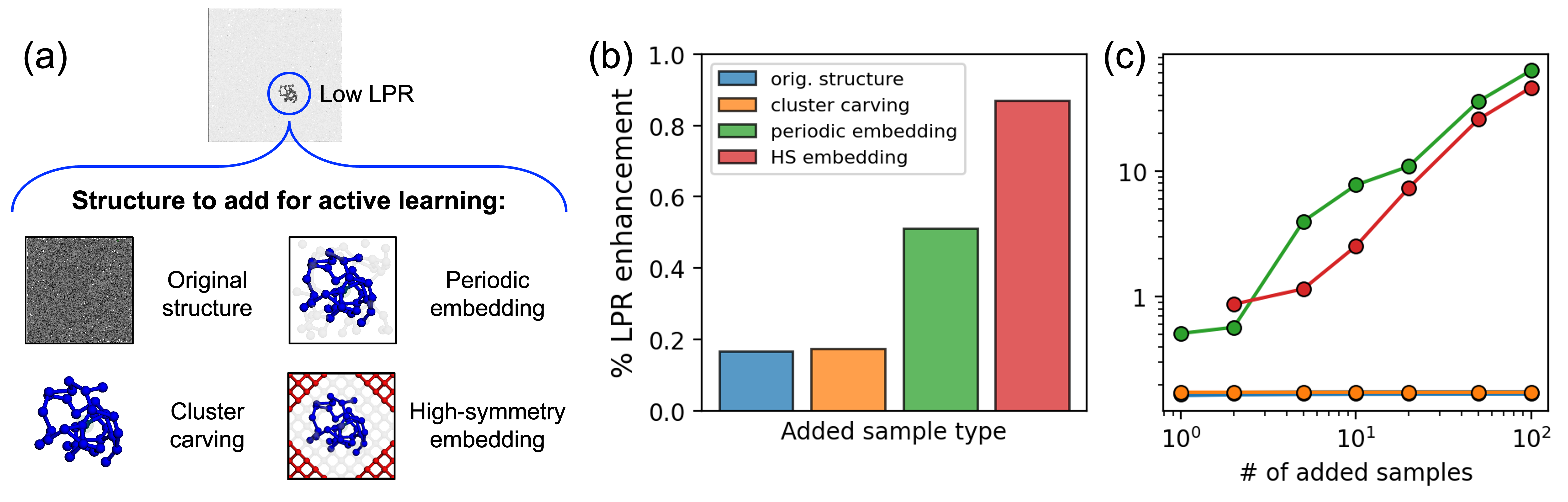}
  \caption{LPR enhancements for the local environment of high uncertainty when different strategies are employed to acquire the additional samples for active learning. Panel (a) is a schematic that visually explains the different strategies of obtaining the additional structures. Blue atoms comprise the identified local environment of high uncertainty. In high-symmetry (HS) embedding, atoms in the HS local environment are shown in red. In both embedding cases, buffer atoms are shown in light gray. Note that the systems are not in scale with one another. Panel (b) presents the LPR enhancements when a \textit{single} additional structure is used, except for the case of HS embedding where the diamond structure used for embedding is also added. Panel (c) shows the LPR enhancements vs. the number of added samples. In the case of original structure inclusion and cluster carving, the same sample is added multiple times.}
  \label{fig:active-learning}
\end{figure*}

Figure \ref{fig:data-aug}a shows that the PR-guided strategy unsurprisingly yields significantly higher PRs for the surface-containing structures. Random selection of low-density bulk structures exhibits higher PRs than complete random selection, but is much less effective. In Figure \ref{fig:data-aug}b, the PR-guided strategy efficiently diminishes the root mean squared error (RMSE) for the surface-containing structures by 0.297 eV with only one additional structure. The RMSE remains low for the proposed strategy from there on, with the lowest RMSE of 0.107 eV observed at four additional structures. Both random selection approaches perform poorly in diminishing the RMSE for the surface-containg structures, exhibiting large fluctuations and higher RMSE than the initial model in some cases. All in all, the PR-guided dataset augmentation strategy successfully identifies a small set of structures that best decrease the error for the target systems, without any explicit model training or reference calculations. The strategy goes beyond simple chemical intuition, as it selects the samples among the low-density ones that are the most adept at reducing the RMSE for the target systems. Our proposed strategy would be even more useful when there does not exist a clear, chemically intuitive approach of selecting the additional structures for fine-tuning.

\subsection{Active learning}

Active learning is a ML technique where the model predictions are continuously assessed during their application, and if the predictions do not satisfy a certain criterion, new training samples are added and the model is re-trained.~\cite{Ang2021, Schwalbe2021, Podryabinkin2022, Lysogorskiy2023, Jalolov2024, Erhard2024, Zaverkin2024} The simplest realization of this strategy based on the PR formalism would be to estimate the uncertainties as the inverse of the PR during inference, identify the samples where the predictions fall below a certain confidence threshold, and add them to the training set to increase the reliability of the model. In Appendix \ref{app:increase-by-one}, we show that this approach is guaranteed to achieve the desired effect, since adding a sample to the training set results in an increase of its PR by one.

In atomistic ML, however, active learning is often employed at the \textit{local} level to identify the environments for which the model exhibits high uncertainty, and add new samples that best improve the model accuracy for those environments. Here, the simplest approach may still be to directly add the structure that contains the local environment of high uncertainty. Nevertheless, this is not always possible: in cases where the ML models are used to simulate large bulk chemical systems, reference calculations for the problematic structures of interest would be prohibitive. An alternative approach is to exploit the locality of atomistic ML models and add smaller structures that still contains the local environment of high uncertainty. This then raises the question of what would be the best approach to obtain the smaller representative structures. In this subsection, we use the LPR to assess different approaches of obtaining the small representative structures for active learning.

For the case study, we use a linear LE-ACE model~\cite{Bigi2022} trained to predict the total energies of 500 randomly selected carbon structures from the entire GAP-17 dataset.~\cite{Deringer2017} 
We then consider performing a single active learning iteration for the model, targeting a liquid carbon structure with 13,824 atoms from a large-scale molecular dynamics (MD) simulation. Further details regarding model training and acquisition of the MD structure are given in the Supplementary Information. After identifying the local environment with the highest uncertainty (lowest LPR) in the large structure, we investigate the following strategies of small representative structure construction (see Figure \ref{fig:active-learning}a for illustrations):
\begin{itemize}
    \item Cluster carving~\cite{Podryabinkin2022}: the local environment is simply treated as a non-periodic spherical cluster with the high uncertainty atom at the center
    \item Periodic embedding~\cite{Lysogorskiy2023, Jalolov2024, Erhard2024}: a cube tightly containing the local environment of interest is extracted to generate a smaller periodic system
    \item High-symmetry (HS) embedding: the local spherical environment is removed from the original structure and embedded into a high-symmetry, crystalline structure
\end{itemize}
In the case of periodic embedding, the unit cell dimensions are adjusted so that close-contact distances between atoms are above 1 Å to avoid non-physical atomic configurations. In the HS embedding, the unit cell dimensions of the HS structure are expanded to a minimum size that includes both the local environment of interest and a local environment from the high-symmetry structure, whilst satisfying the close-contact criterion. Exceptionally, in this case, the HS structure used for embedding (diamond in our case study) is also added to the training set. In both embedding approaches, ``buffer atoms'' that exist outside of the fixed local environments are randomly displaced by sampling from a Gaussian distribution with $\sigma = 0.2$ Å. Apart from the listed strategies, inclusion of the entire target structure is also considered as a baseline.

In Figure~\ref{fig:active-learning}b, both original structure inclusion and cluster carving exhibit LPR enhancements slightly below 0.2\%.
From the LPR perspective, since the original structure contains a large number of atoms, there is large, unresolved arbitrariness in how the total energy of the structure is partitioned into the local contributions.
In the case of cluster carving, lack of information on the local environments other than the one of interest leads to the unresolved arbitrariness, especially given that the other local environments encompass the cluster surface that most likely does not appear in the original training set.
As a result, the LPR enhancements for these two approaches are rather low. 

The two embedding strategies result in comparably larger LPR enhancements of 0.5\% for periodic embedding and 0.8\% for HS embedding. In the periodic embedding case, using a much smaller unit cell tightly bound to the local environment of interest results in a smaller number of atoms, and this allows the model to better resolve the uncertainty in the local environment of interest. In the case of HS embedding, the strategy benefits from similar factors as well as the coexistence of a local environment from the HS structure, diamond. By additionally including the diamond structure, the LPR for the diamond environment is fully resolved,~\cite{Chong2023} and hence the LPR of the target local environment gets further enhanced.

\vspace{0.25cm}
Results in Figure \ref{fig:active-learning}c show that inclusion of multiple structures can further resolve the uncertainty in the target environment. In the case of periodic embedding, inclusion of 10 structures as opposed to 1 increases the LPR enhancement from 0.5\% to 7.7\%. 
This is explained by the presence of buffer atoms and their random displacement between multiple structures, which effectively resolves their local degeneracies. For this particular case study, the LPR enhancement for HS embedding becomes lower than periodic embedding with a few more added samples due to a larger number of buffer atoms (see Figure \ref{fig:active-learning}a). These results reveal the clear benefits of adopting an embedding approach and adding \textit{multiple} structures at each iteration of the active learning process to best resolve the local uncertainties encountered by atomistic ML models. Note that first-principles calculations for 10 small representative structures would still be immensely cheaper than that for the original structure. In practice, a threshold can be implemented to ensure that enough structures are used to sufficiently resolve the uncertainties in the identified local environment at each active learning iteration. Lastly, one can envision more realistic methods of imposing the displacements on the buffer atoms (e.g. constrained MD ~\cite{Erhard2024}).

\newpage

 \begin{figure*}[t!]
\centering
  \includegraphics[width=17cm]{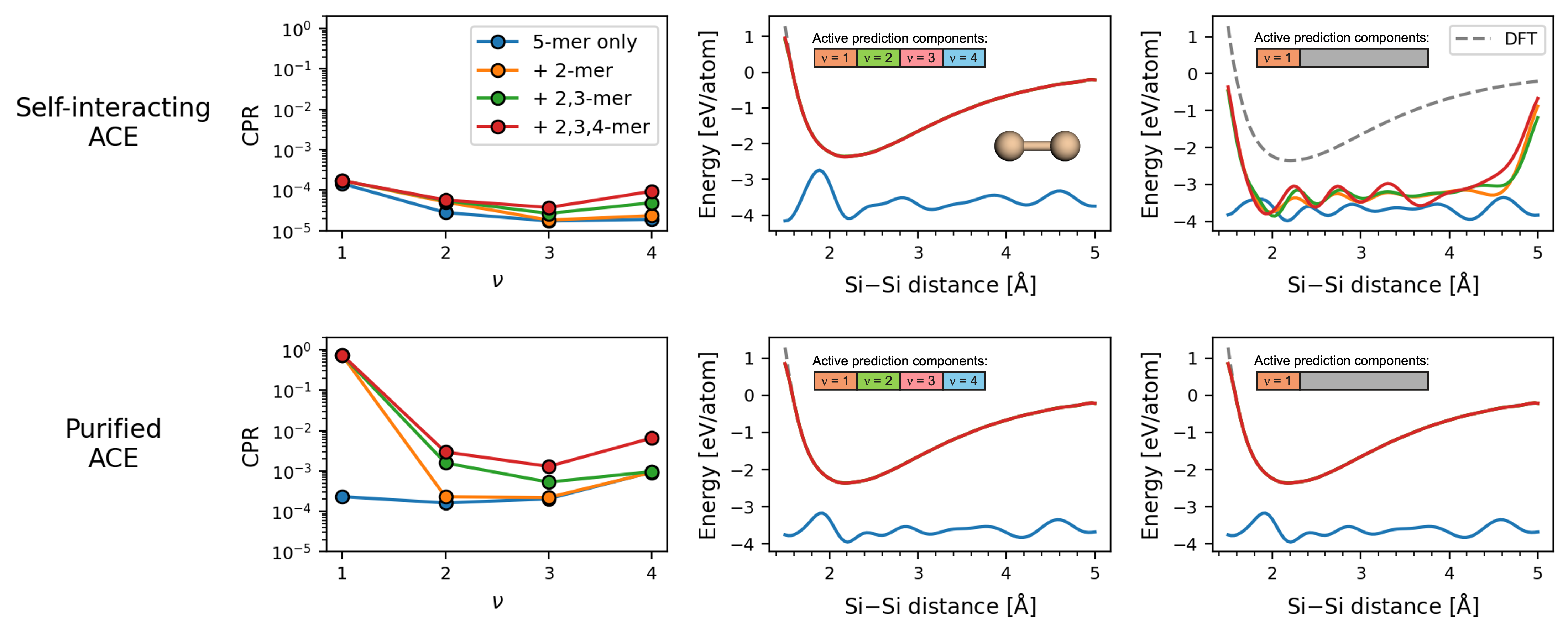}
  \caption{Difference in the learning behavior of linear ACE models before and after purification of the feature vectors. Results for the self-interacting ACE are shown in the top row, and for the purified ACE are shown in the bottom row. The left column shows the differences in the CPR for a test set of silicon pentamers as the original training set of silicon pentamers is amended to include dimer, trimer, and tetramers, successively. The middle column shows the predicted energies along the Si--Si dimer curve, when the entire feature vector is used for energy prediction. The right column shows only the the $\nu=1$ (two-body) component of the predicted dimer energies. In the latter two rows, total energy of the dimers from DFT calculation is shown with a dashed gray line for reference.}
  \label{fig:cpr-body-order}
\end{figure*}

\section{Component-wise prediction rigidity}\label{sec:cpr}

In many cases, a single type of descriptor falls short in adequately describing the system of interest. Atomistic ML models are hence often constructed by using a concatenation of multiple sets of features that describe the same system in different ways. A straightforward example is given by body-ordered expansion approaches,~\cite{Drautz2019, Batatia2022BOTNET, Nigam2022, Bigi2023} where the descriptor is a combination of several components that describe the same local environment in terms of increasingly large groups of neighbors. Another example is the use of a long-range descriptor in conjunction with a short-range descriptor to allow the ML model to learn the chemical system at multiple length scales.~\cite{Grisafi2019, Grisafi2021, Huguenin2023}

In common model architectures, these different components in the descriptors contribute to the prediction separately. That is, the global prediction of the model is expressed as a sum over the prediction components, which shares a resemblance with classical force fields.
In this case, one can compute the component-wise prediction rigidity, or the CPR, for the individual prediction components (see Table \ref{tbl:prs_extended}).
The CPR then allows one to diagnose the model by considering the prediction components individually, allowing for a more practical understanding of where the model succeeds or fails, which model component needs improvement, and whether the decomposition is robust and hence interpretable.

For the remainder of this section, we consider the CPR of a linear atomic cluster expansion (ACE) model~\cite{Drautz2019} as well as a multi-length-scale model that combines SOAP~\cite{Bartok2013} and long-distance equivariant (LODE)~\cite{Grisafi2019} descriptors. In both cases, we first use the CPR to expose the non-orthogonality of conventional approaches in computing the descriptors and its implication on the learning behavior of the resulting model. We then compute the CPR for the case where the different components of the descriptors are made orthogonal with respect to one another, revealing the clearly distinct learning behavior of the ML models as a result of such treatment.

\subsection{Body-orderedness of linear ACE models}

ACE is a many-body expansion formalism~\cite{Drautz2019} that involves the reformulation of the canonical many-body expansion into another expansion that also includes ``self-interactions'' (i.e. higher body-order contributions where the same atom is counted multiple times),~\cite{Willatt2019} allowing for much greater efficiency in computing the descriptors.
While the success of the ACE formalism is evident from the literature,~\cite{Kovacs2021,Dusson2022,Qamar2023,vanderOord2023} ML models adopting the ACE formalism describe the chemical systems with the spurious self-interactions included. 
Here, we use the CPR to investigate the impact of the self-interactions in the nature of ACE models and its implication on their learning behavior.

We consider linear ACE models~\cite{Witt2023} where the highest correlation order $\nu_{\mathrm{max}}$ is 4. An initial model is first trained on a dataset of 500 randomly generated silicon pentamers, training on their total energies. Next, successively modified training sets are obtained by replacing 50 samples with dimers, then another 50 with trimers, and finally 50 more with tetramers. Separate linear ACE models are then trained on each of the modified datasets. Details of model training and silicon cluster generation are given in the Supplementary Information. Here, one can interpret the models based on the fact that the ACE feature vector is a concatenation of multiple blocks that each correspond to a different $\nu$. Then, since the weights applied on different $\nu$ blocks are strictly independent, each block can be seen as a separate prediction component. One can then individually compute the CPRs for the individual $\nu$ components, as well as their energy contributions. Note that based on this interpretation, the successive inclusion of lower $n$-mers to the training set is aims to resolve the degeneracies in the energy partitioning between the different $\nu$ components.

The top row of Figure \ref{fig:cpr-body-order} shows the CPRs and energy predictions of conventional, self-interacting ACE models. The CPRs remain low across all $\nu$ components for the four models considered, with no resolution taking place as the lower $n$-mers are added to the training set. In the total energy predictions for silicon dimers, the three models that have seen the dimer configurations during training are able to recover the reference dimer curve accurately. However, the $\nu = 1$ component of the energy has no resemblance to the dimer energy in all four cases, which is a clear indication of the arbitrary partitioning reflected by the low CPR.

Recently, Ho et al.~\cite{Ho2024} have introduced a ``purification'' operator for ACE, which eliminates the self-interactions and allows for the exclusive consideration of canonical contributions to the many-body expansion in the computation of ACE features. To investigate the effect of purification on the learning behavior of ACE models, the above exercise was repeated for the purified ACE models. In the bottom row of Figure \ref{fig:cpr-body-order}, one sees that the purified ACE models are capable of resolving the partitioning degeneracy between different $\nu$ components, as evident from the significant increase in the CPR of the $\nu$ component when samples of the corresponding $n$-mers are included in the training set. Interestingly, the final addition of tetramers also leads to a notable increase in the CPR for $\nu = 4$, which corresponds to the pentamers. This is a sign that the degeneracy across all of the $\nu$ components has been largely resolved. Such a trend in the CPR is reflected by a distinct behavior in the energy predictions: in the case of purified ACE models, both the total energy predictions and the $\nu = 1$ energy components are capable of recovering the reference dimer curve.

These results altogether reveal that the matching of the $\nu$ feature blocks with their respective body-orders is not possible in the presence of self-interaction terms in conventional ACE models. As an example, the models learn the dimer energetics by using not only the $\nu = 1$, but also all of the other $\nu$ components. It is only with purification, which removes the spurious self-interaction terms, that the ACE models become capable of learning in an explicitly body-ordered manner.

\begin{figure*}[t!]
\centering
  \includegraphics[width=14cm]{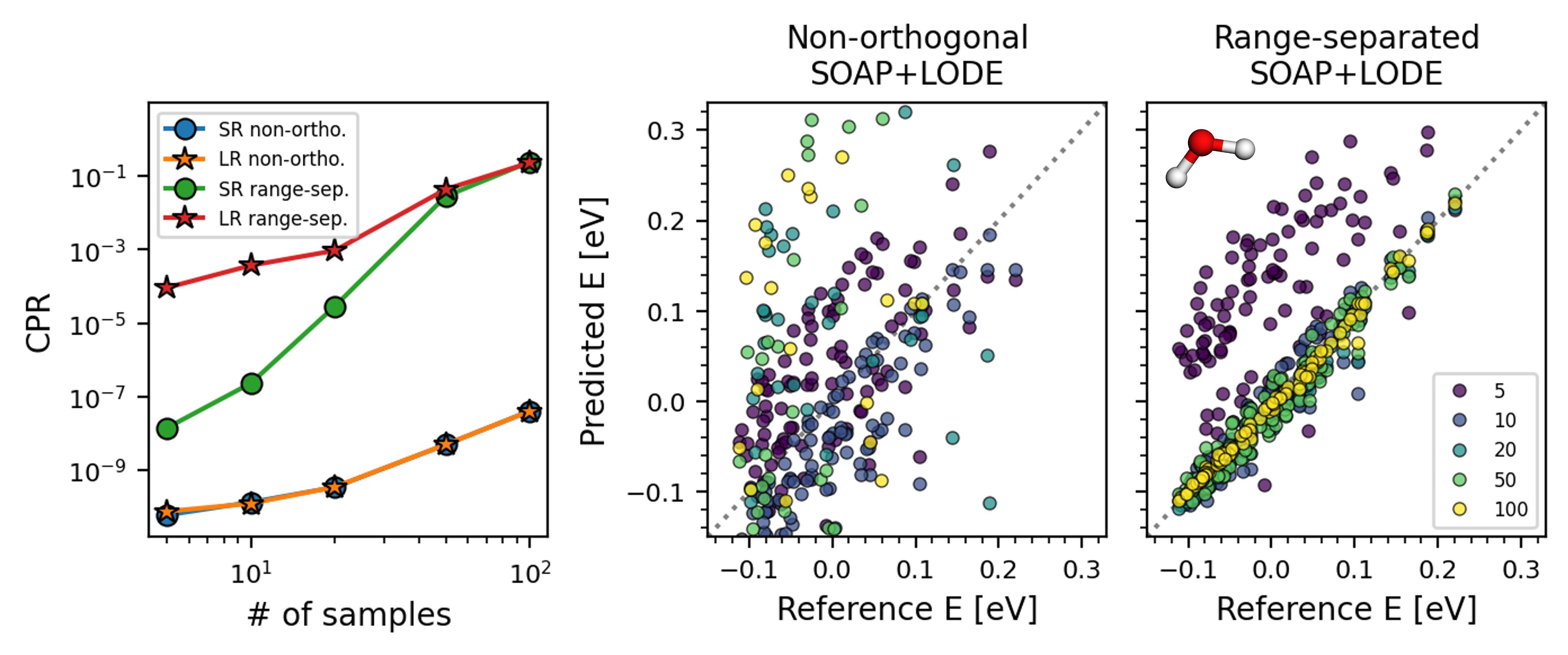}
  \caption{Differences in the learning behavior of SOAP+LODE models trained on water dimers before and after range separation. The left panel shows the CPRs of the models for the short-range (SR) and long-range (LR) prediction components. The right panels show the extrapolative energy predictions on water monomers made by the two models. Note that in the case of non-orthogonal SOAP+LODE, not all data points are present within the displayed energy range. Zero of the energies are set to the monomer dataset mean.}
  \label{fig:cpr-sr-lr}
\end{figure*}

\subsection{Range separation in multi-length-scale model}

Many atomistic ML models employ a locality ansatz where the global property is expressed as a sum of local contributions, based on the nearsightedness principle of electronic matter.~\cite{Prodan2005}
This means that the models are incapable of incorporating structural information beyond a fixed radius around the central atom.
While such a description is sufficient in many cases, there exist several instances where it cannot be, most notably when long-range physics is present within the target system.
To overcome this deficiency, several strategies have been proposed,~\cite{Morawietz2012, Bereau2015, Yao2018, Unke2019} one of which is the use of long-range atomic descriptors such as LODE.~\cite{Grisafi2019, Grisafi2021, Huguenin2023}
LODE replaces the the Gaussian or delta functions placed on the atoms with Coulomb potentials that possess $1/r^p$ tails. This allows the model to account for the long-range interactions while retaining the atom-centered approach in describing the chemical systems.
In practice, LODE is often used in conjunction with a short-range descriptor, such as SOAP, to allow the ML models to account for multiple length scales.

Here, we investigate the differences in the ML model learning behavior before and after strict range separation, i.e. eliminating any double counting of atoms between prediction components that correspond to different length scales.
To this end, two distinct implementations of SOAP+LODE models are considered: in the first case of non-orthogonal SOAP+LODE, the LODE descriptor is simply computed in reciprocal space, accounting for the contributions from all atoms in the periodic system. This results in a double counting of the atoms within the short-range (SR) cutoff set at 2.8 Å, where they contribute to both SOAP and LODE descriptors. In the second case of range-separated SOAP+LODE, the abovementioned LODE descriptor is further treated by subtracting the contributions from the atoms within the SR cutoff.
A dataset composed of 100 water dimer configurations and their total energies is used for model training. In all configurations, dimers are separated by more than 3 Å so that only the long-range (LR) prediction component of the model can capture the intermolecular interactions.
Then, to understand the effects of range separation, extrapolative energy predictions of the models on 100 monomer configurations is considered. Further details on the model training and dataset construction are provided in the Supplementary Information.

The left panel of Figure \ref{fig:cpr-sr-lr} shows changes in the CPR with respect to the training set size.
In the case of non-orthogonal SOAP+LODE, both the SR and LR prediction components exhibit low CPR values throughout all considered training set sizes, and the values remain very close to one another.
On the contrary, range-separated SOAP+LODE shows a marked difference in the CPR at lower number of training samples, with a higher CPR for the LR component. As the training set size grows, the two converge to a high CPR value, corresponding to a difference of seven orders of magnitude when compared to the non-orthogonal case.
The water monomer energy prediction results reveal that the non-orthogonal SOAP+LODE models extrapolate poorly to the monomers, yielding worse predictions as the training set size grows. Conversely, the extrapolative performance of the range-separated SOAP+LODE models improve with the training set size, and the final model is able to make accurate predictions with an RMSE of 7.6 meV (as opposed to 939 meV of non-orthogonal SOAP+LODE).

The results here can be explained as the resolution in the degeneracy between the SR and LR components of range-separated SOAP+LODE that takes place as more samples are added.
Since the water dimer dataset spans a range of different separation distances from 3 to 10 Å, when the separation is large, the LODE block of the range-separated SOAP+LODE descriptor converges to zero, allowing SOAP to obtain an accurate description of the individual monomers.
Such effects are promptly captured by the different trends in the CPR observed for the non-orthogonal and range-separated SOAP+LODE models.

Both case studies presented in this section demonstrate that without carefully considering the overlap between different prediction components, ML models may utilize the available features in an unexpected manner, where multiple prediction components are used to learn the physics that can be sufficiently described by only one. While feature orthogonalization does not guarantee a significant improvement in accuracy,~\cite{Ho2024} one should still recognize the benefits in ensuring that each prediction component is used for its originally engineered purpose. The CPR provides an easy strategy to individually gauge the robustness of intermediate predictions made by the model.

\begin{figure*}[t!]
\centering
  \includegraphics[width=16cm]{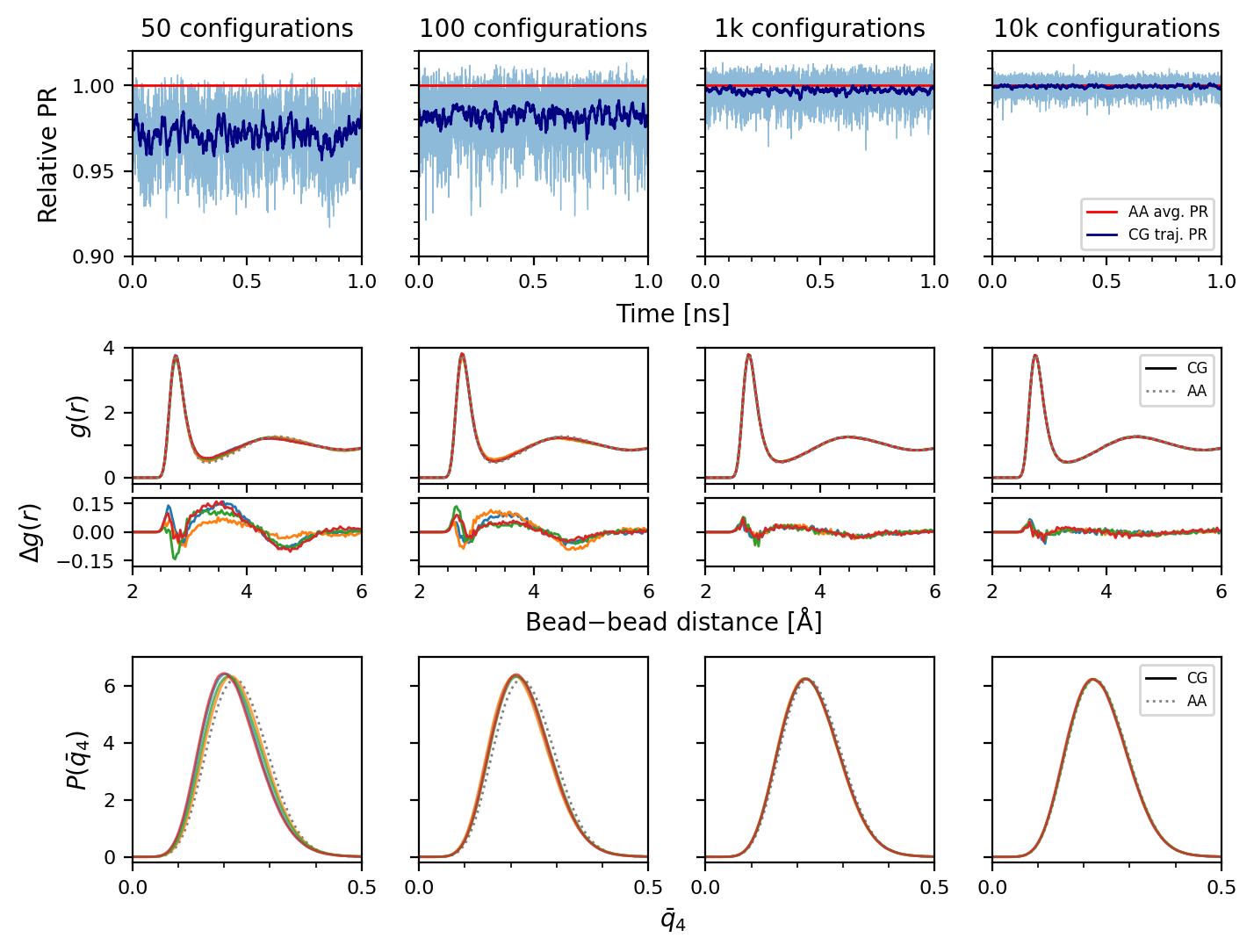}
  \caption{Coarse-grained water MD simulation results of the trained MACE models. The top row shows the PR along a single MD trajectory normalized by the average PR for the test set. The light line shows the raw uncertainty values, and the dark line shows the moving average over 200 timesteps. Red line marks the average energy uncertainty of the test set computed for each model. The middle row shows the pair correlation function $g(r)$, and the bottom row shows the average $l = 4$ local Steinhardt order parameter $\bar{q}_4$ distribution, computed for four coarse-grained MD runs of different models trained on randomly sampled data. The reference distribution of the all-atoms simulation is also shown in both rows.}
  \label{fig:cg-water}
\end{figure*}

\section{Application to coarse-grained ML}\label{sec:cg}

In the computational simulation of proteins and other macromolecules, coarse-graining techniques are often employed to study the system of interest at a significantly reduced cost by combining rigid and/or unreactive groups of atoms into pseudo-atoms, or beads, and sampling their configurations through an effective potential built to match the statistical behavior of the all-atoms simulation.~\cite{Monticelli2008, Kmiecik2016, Souza2020} 
Recently, such approaches have also been combined with ML interatomic potentials, allowing researchers to benefit from the highly versatile functional forms offered by the ML techniques in studying the large-scale systems of interest.~\cite{Wang2019, Durumeric2023, Sahrmann2023}

Here, one should note that the conventional coarse-graining approach of ``force-matching''~\cite{Izvekov2004, Noid2008} leads to the absence of an explicit energy target for model training. The quality of the model can then be ascertained by verifying that that thermodynamic properties, such as configurational distributions, match those of an all-atoms simulation. Another aspect to recognize is the large noise from the non-bijective relationship between all-atomic and coarse-grained systems. As multiple all-atomic configurations with different energetics can be represented by the same coarse-grained configuration, a large noise is expected to be present in the reference data, and the ML models are expected to learn the underlying ``potential of mean force'' (PMF).~\cite{Wang2019} Given these complications, it is ever more crucial to devise methods that can reliably provide the uncertainties associated with the predictions of ML models for coarse-grained systems.~\cite{Duschatko2024}

In this final section, we demonstrate the applicability of the PR formalism for MACE models trained on coarse-grained water, a system explored by several others in previous ML studies.~\cite{Zhang2018, Chan2019}
To generate the training data, we performed classical all-atoms MD simulation for an NVT ensemble at 300 K.
The trajectories are coarse-grained by taking the center of mass of each water molecule as the bead position, and separately summing the force components of the constituent atoms to compute the three force components of each bead.
The MACE models are trained on training set sizes of 50, 100, 1000, and 10,000 configurations, where each configuration contains 128 beads that provide 384 force targets in total. For each training set size, random sampling and model training is repeated four times. Fixed validation and test sets of 1000 configurations each are used.
The trained models are used to run 1 ns simulations of a 128-bead coarse-grained water system under the same conditions as the reference all-atoms simulation.
The accuracy of the models is considered by calculating the relative PR of the resulting trajectories with respect to the test set average, as well as comparing the pair correlation function, $g(r)$, for two-body correlations,  and the average $l=4$ local Steinhardt order parameter $\bar{q}_4$ distributions~\cite{Steinhardt1983, Zhang2018} for higher body-order correlations. Full details of dataset generation, model training, and MD simulations are provided in the Supplementary Information.

In the top row of Figure \ref{fig:cg-water}, the MACE models exhibit different degrees of deviation in the relative PR from the reference data for the different training set sizes.
As the training set size increases, the models better distinguish and learn the underlying PMF, and as a result, PRs for the simulated system trajectory converges to that observed for the test set configurations.
A similar trend is also captured in the $g(r)$ and $\bar{q}_4$ distributions presented in the middle and bottom row of Figure \ref{fig:cg-water}, respectively. At 50 training configurations, both $g(r)$ and $\bar{q}_4$ distributions of the coarse-grained model MD trajectories notably deviate away from the reference distribution. As the training set size increases, they get closer and closer to the reference, until at 10,000 configurations, good agreement between the coarse-grained MD trajectories and the reference is observed. These results reveal that the PRs are useful in assessing the robustness of coarse-grained ML model predictions and tracking their training convergence. 
For this application, the PR provides only a qualitative indicator of convergence, as it is not possible to convert it into a calibrated uncertainty estimate, given the intrinsic error in the forces associated with the coarse-graining procedure. Still, it can be very useful, and more informative than the validation force error that saturates quickly to the limiting coarse-graining error: when going from 50 to 10,000 training configurations, it drops only slighly from 150 to 145 meV/Å.
In Appendix~\ref{app:cg-lpr}, we further show that the LPR of the models can validly detect the \emph{local} uncertainty along a MD trajectory.

\section{Conclusion and outlook}

Throughout this work, we have established the PRs as a highly versatile set of tools to understand, and enhance, the robustness of ML model predictions, presenting many concrete examples for data-driven chemistry. 
We have shown that the PRs can quantify the robustness of local and global predictions for various NN architectures. We then revealed that the PR distribution for a NN model with a fixed architecture shows dependence on dataset makeup while being largely insensitive to the optimization details.
Next, we have presented the utility of the PRs in guiding target-oriented dataset augmentation and active learning, where the metrics can be used to identify a set of structures that can best reduce the error for a target system or resolve the uncertainty for a local environment.

We have also extended the PR formalism to the case where the model predictions are made as a sum over several prediction components. There, our metrics uncovered that without proper orthogonalization of the features, model learning behavior can deviate significantly from expectations, and that, for example, commonly adopted body-ordered or range-separated architectures cannot be interpreted in terms of clearly-separated contributions.
Finally, we have demonstrated the wide applicability of the PRs by applying the formalism to NN models for coarse-grained water and showing that the PRs correlate well with the accuracies in the macroscopic observables from the MD simulations performed with the trained models.

The underlying mathematical formulation for PRs can be applicable to a wide variety of ML models, even those trained on experimental reference data. It is, however, presently limited to regression models where the prediction is made linearly with respect to the (last-layer) features. Future research efforts should extend the formalism to models with non-linearity in the prediction layer, such as classification models. All in all, the PRs are ideal metrics to adopt for improving the interpretability and transferability of data-driven techniques, which we hope will contribute to reliable machine learning practices in the field of chemical sciences.
\newline
\section*{Author Contributions}

S.C., F.B., F.G., and M.C. conceptualized the prediction rigidity formalism. S.C. and F.G. designed the computational experiments, and S.C., F.B., P.L., and M.K. carried out the computational experiments. M.C. oversaw the research efforts and acquired funding. All authors participated in data analysis and writing of the manuscript.

\section*{Conflicts of interest}

There are no conflicts of interest to declare.

\section*{Acknowledgments}

The authors would like to thank F\'elix Musil and Guillaume Fraux for the insightful discussions and help with code implementation. Christoph Ortner and Cheuk Hin Ho are gratefully acknowledged for discussions and sharing of an early version of the ACE purification code. S.C. and M.C. acknowledge the support by the Swiss National Science Foundation (Project 200020\_214879). F.B., P.L., and M.C. acknowledge support from NCCR--MARVEL, funded by the Swiss National Science Foundation (grant no. 182892). F.G., and M.C. acknowledge funding from the European Research Council (ERC) under the European Union’s Horizon 2020 research and innovation programme (grant no. 101001890--FIAMMA).

\section*{Data availability}

All of the datasets, codes, and scripts used in this study can be accessed at \url{https://github.com/SanggyuChong/faraday_discussions_2024} and on Materials Cloud.~\cite{Talirz2020}

\bibliographystyle{rsc}  

\bibliography{biblio}

\appendix

\section{Prediction rigidities for generic loss functions}\label{app:generic-loss}

Although the main text focuses on the ubiquitous case of a sum of squared error loss, it should be noted that prediction rigidities can be calculated for any loss functions. In any case, Computing $\mathbf{H}_o$ can be prohibitively expensive and requires an implementation of the second derivatives of the model. Therefore, an approximation of the Hessian is often used in practice, where
\begin{equation}\label{eq:pseudo-hessian}
     \mathbf{H} \approx \sum_{i \in \mathcal{D}} \Big( \frac{\partial y_i}{\partial \mathbf{w}} \Big)^\top \frac{\partial^2 \ell_i}{\partial \mathbf{w} \partial \mathbf{w}^\top} \,\, \frac{\partial y_i}{\partial \mathbf{w}}.
\end{equation}
This pseudo-Hessian, also known as generalized Gauss-Newton Hessian, does not contain any second derivatives of the model, and it is equivalent to the full Hessian in the important case of a linear model trained with a loss function corresponding to the sum of squared errors. We recommend this formulation for most application of the PRs.

\section{Proof that PR for a sample increases by one upon its addition to the training set}\label{app:increase-by-one}

Assume a loss given by a sum of squared errors. We rename, without any loss of generality,
\begin{equation*}
    \frac{\partial y_i}{\partial \mathbf{w}} = \mathbf{x}_i,
\end{equation*}
where $\mathbf{X}$ is the matrix that stacks all the $(\partial y_i / \partial \mathbf{w})^\top$ and $i \in \mathcal{D}$. The PR before the addition of the structure $\star$ is given by
\begin{equation*}
    b^{-1} = (\mathbf{x}_\star^\top (\mathbf{X}^\top \mathbf{X})^{-1} \mathbf{x}_\star)^{-1},
\end{equation*}
while the PR after the addition of the addition of the structure is
\begin{equation*}
    a^{-1} = (\mathbf{x}_\star^\top (\mathbf{X}^\top \mathbf{X} + \mathbf{x}_\star \mathbf{x}_\star^\top)^{-1} \mathbf{x}_\star)^{-1}.
\end{equation*}
The difference between the two is therefore
\begin{equation*} 
\begin{split}
    &a^{-1} - b^{-1} = a^{-1}(a-b)b^{-1} \\ &=
    a^{-1}(\mathbf{x}_\star^\top (\mathbf{X}^\top \mathbf{X} + \mathbf{x}_\star \mathbf{x}_\star^\top)^{-1} \mathbf{x}_\star-\mathbf{x}_\star^\top (\mathbf{X}^\top \mathbf{X})^{-1} \mathbf{x}_\star)b^{-1} \\ &=
    a^{-1}(\mathbf{x}_\star^\top ((\mathbf{X}^\top \mathbf{X} + \mathbf{x}_\star \mathbf{x}_\star^\top)^{-1}-(\mathbf{X}^\top \mathbf{X})^{-1}) \mathbf{x}_\star)b^{-1} \\ &=
    a^{-1}(\mathbf{x}_\star^\top (\mathbf{X}^\top \mathbf{X} + \mathbf{x}_\star \mathbf{x}_\star^\top)^{-1} \mathbf{x}_\star \mathbf{x}_\star^\top (\mathbf{X}^\top \mathbf{X})^{-1} \mathbf{x}_\star)b^{-1} \\ &=
    a^{-1}abb^{-1} = 1,
\end{split}
\end{equation*}
where the matrix identity $\mathbf{A}^{-1} - \mathbf{B}^{-1} = \mathbf{A}^{-1}(\mathbf{A}-\mathbf{B})\mathbf{B}^{-1}$ has been used twice (the first of which is applied on the two scalars $a$ and $b$). Trivially, this proof also holds in the case where the same regularization term is added to $\mathbf{X}^\top \mathbf{X}$ and $\mathbf{X}^\top \mathbf{X} + \mathbf{x}_\star \mathbf{x}^\top_\star$ before the inversion.

\section{Utility of LPR in coarse-grained ML}\label{app:cg-lpr}

\begin{figure}[b!]
\vspace{2cm}
\centering
\includegraphics[width=\columnwidth]{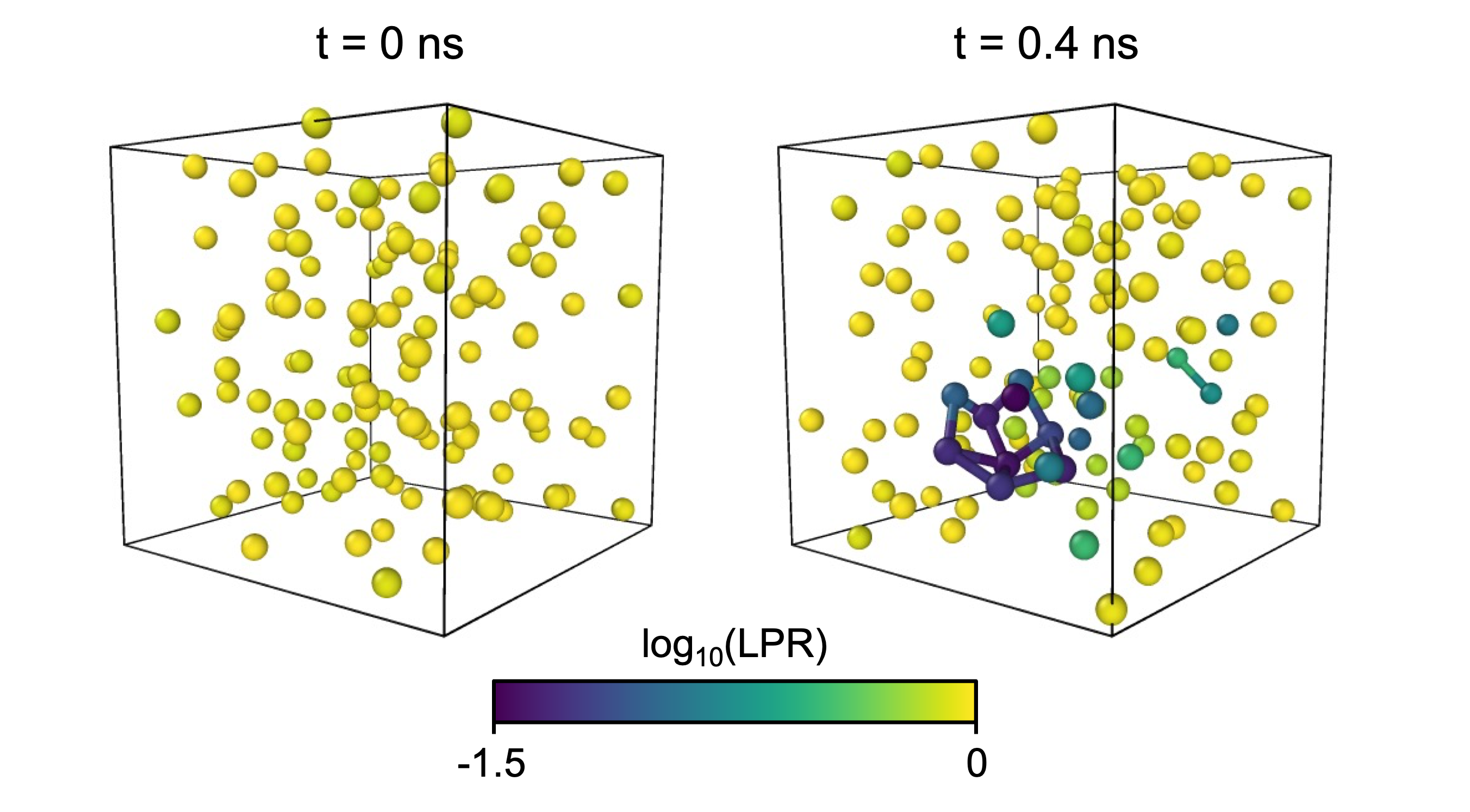}
  \caption{Bead configurations from a MD trajectory that becomes energetically unstable from $t = 0.4$ ns. Each bead corresponds to a single water molecule. The initial configuration is shown on the left, and the first energetically unstable configuration is shown on the right. Beads are colored by their LPRs. The LPR values are normalized to the mean of the initial configuration. In both cases, interatomic distances smaller than 2.5 Å are expressed as bonds.}
  \label{fig:cg-lpr}
\end{figure}

In the ML models for coarse-grained systems, the LPR quantifies the robustness of local predictions made for individual coarse-grained \textit{beads} in the system. To showcase its utility, an energetically unstable MD trajectory is obtained with the MACE model trained on 1000 reference configurations from Section~\ref{sec:cg}, and the LPRs are computed for the initial configuration and the configuration at which energy instability is first observed. In Figure \ref{fig:cg-lpr}, one can see that the LPR distribution for the initial configuration is relatively uniform without any outliers. In the problematic configuration, the LPR values are lower for the beads that are in close contact with one another ($< 2.5$ Å), which the model has never seen during training and are hence the sources of energy instability. Apart from the beads that are involved in the close-contact network, several other beads are also observed with notably lower LPR values, which indicates that the LPR is capable of detecting local uncertainties beyond what can be deduced from simple chemical intuitions.

\end{document}